\documentclass[sigconf]{acmart}
\usepackage{enumitem}
\usepackage[hyphenbreaks]{breakurl}
\usepackage{multirow}
\usepackage{cleveref}
\usepackage{balance}
\usepackage{threeparttable}
\newcommand\Tstrut{\rule{0pt}{3ex}}         % = `top' strut
\newcommand\Bstrut{\rule[-2ex]{0pt}{0pt}}   % = `bottom' strut

\def\BibTeX{{\rm B\kern-.05em{\sc i\kern-.025em b}\kern-.08emT\kern-.1667em\lower.7ex\hbox{E}\kern-.125emX}}

\copyrightyear{2019}
\acmYear{2019}
\setcopyright{acmlicensed}
\acmConference[WebSci '19]{WebSci '19: ACM Conference on Web Science}{June 30--July 03, 2019}{Boston, MA}
\acmBooktitle{WebSci '19: ACM Conference on Web Science, June 30--July 03, 2019, Boston, MA}
\setcopyright{none}

\RequirePackage[textsize=scriptsize,backgroundcolor=yellow!30]{todonotes} 
\begin{document}

%
% The "title" command has an optional parameter, allowing the author to define a "short title" to be used in page headers.
\title{What sets Verified Users apart? \\ Insights, Analysis and Prediction of Verified Users on Twitter}

%
% The "author" command and its associated commands are used to define the authors and their affiliations.
% Of note is the shared affiliation of the first two authors, and the "authornote" and "authornotemark" commands
% used to denote shared contribution to the research.
\author{Indraneil Paul}
\affiliation{%
 \institution{IIIT Hyderabad}}
 \email{indraneil.paul@research.iiit.ac.in}
 
\author{Abhinav Khattar}
\affiliation{%
  \institution{IIIT Delhi}}
\email{abhinav15120@iiitd.ac.in}

\author{Shaan Chopra}
\affiliation{\institution{IIIT Delhi}}
\email{shaan15090@iiitd.ac.in}

\author{Ponnurangam Kumaraguru}
\affiliation{%
  \institution{IIIT Delhi}}
\email{pk@iiitd.ac.in}

\author{Manish Gupta}
\affiliation{\institution{Microsoft India}}
\email{gmanish@microsoft.com}

%
% By default, the full list of authors will be used in the page headers. Often, this list is too long, and will overlap
% other information printed in the page headers. This command allows the author to define a more concise list
% of authors' names for this purpose.
\renewcommand{\shortauthors}{Paul, et al.}
\renewcommand{\shorttitle}{Insights, Analysis and Prediction of Verified Users on Twitter}

%
% The abstract is a short summary of the work to be presented in the article.
\begin{abstract}
 Social network and publishing platforms, such as Twitter, support the concept of a secret proprietary \textit{verification} process, for handles they deem worthy of platform-wide public interest. In line with significant prior work which suggests that possessing such a status symbolizes enhanced credibility in the eyes of the platform audience, a verified badge is clearly coveted among public figures and brands. What are less obvious are the inner workings of the verification process and what being verified represents. This lack of clarity, coupled with the flak that Twitter received by extending aforementioned status to political extremists in 2017, backed Twitter into publicly admitting that the process and what the status represented needed to be rethought.
 
 With this in mind, we seek to unravel the aspects of a user's profile which likely engender or preclude verification. The aim of the paper is two-fold: First, we test if discerning the verification status of a handle from profile metadata and content features is feasible. Second, we unravel the features which have the greatest bearing on a handle's verification status. We collected a dataset consisting of profile metadata of all 231,235 verified English-speaking users (as of July 2018), a control sample of 175,930 non-verified English-speaking users and all their 494 million tweets over a one year collection period. Our proposed models are able to reliably identify verification status (Area under curve AUC > 99\%). We show that number of public list memberships, presence of neutral sentiment in tweets and an authoritative language style are the most pertinent predictors of verification status.
 
 To the best of our knowledge, this work represents the first attempt at discerning and classifying verification worthy users on Twitter.
  
\end{abstract}

%
% The code below is generated by the tool at http://dl.acm.org/ccs.cfm.
% Please copy and paste the code instead of the example below.
%



\maketitle

\section{Introduction}
The increased relevance of social media in our daily life has been accompanied by an exigent demand for a means to affirm the authenticity and authority of content sources. This challenge becomes even more apparent during the dissemination of real-time or breaking news, whose arrival on such platforms often precedes eventual traditional media reportage~\cite{b2,b3}. In line with this need, major social networks such as Twitter, Facebook and Instagram have incorporated a verification process to authenticate handles they deem important enough to be worth impersonating. Usually conferred to accounts of well-known public personalities and businesses, \textit{verified accounts}\footnote{The exact term varies by platform, with other social networks using the term ``Verified Profiles''. However in the interest of consistency, all owner-authenticated accounts are referred to as \textit{verified accounts}, and  their owners as \textit{verified users}.} are indicated with a badge next to the screen name (e.g., \scalerel*{\includegraphics{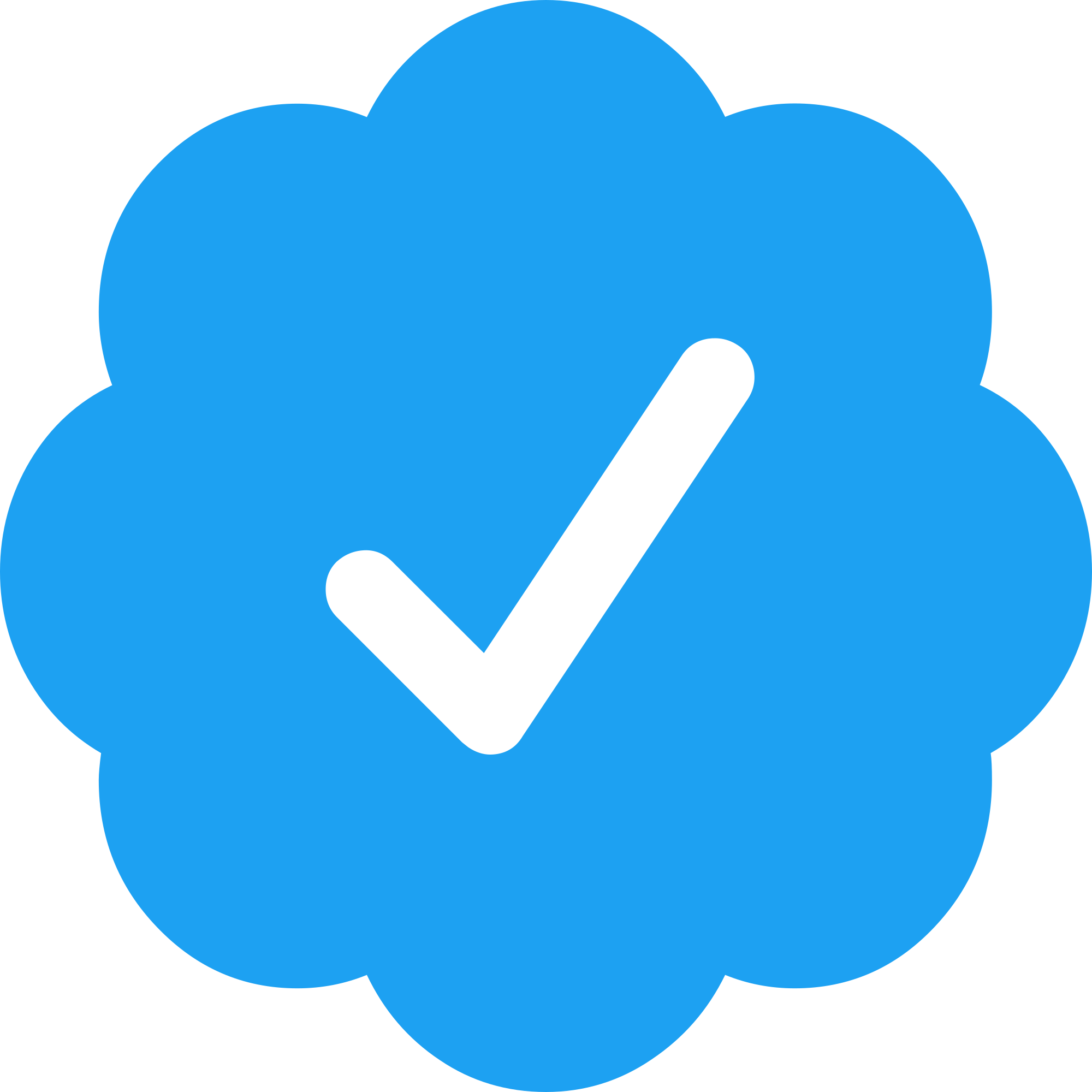}}{B} on Twitter and  \scalerel*{\includegraphics{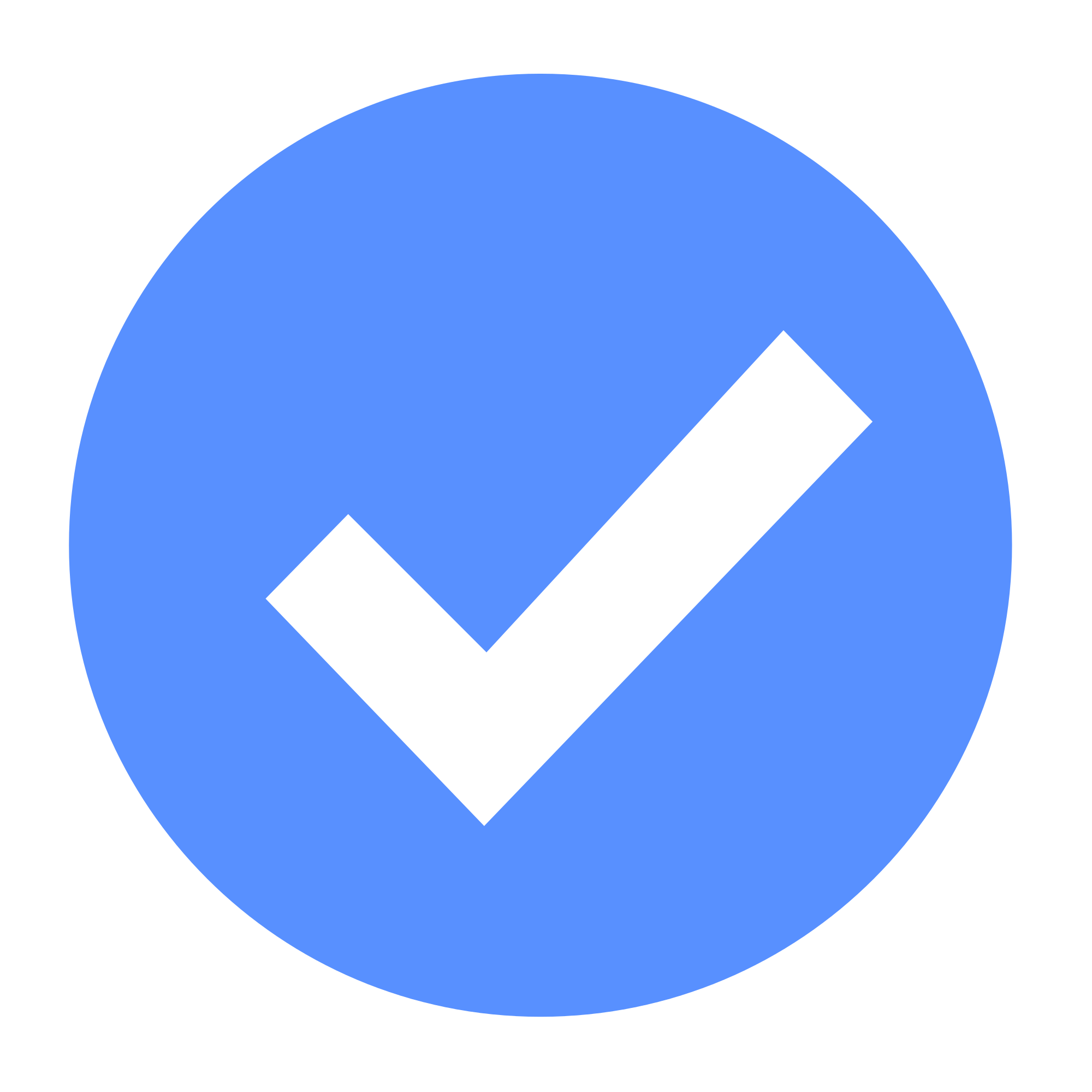}}{B} on Facebook). Twitter's verification policy~\cite{b1} states that an account is verified if it belongs to a personality or business deemed to be of sufficient public interest in diverse fields, such as journalism, politics, sports, etc. However, the exact decision making process behind evaluating the strength of a user's case for verification remains a trade secret. This work attempts to unravel the likely factors that strengthen a user's case for verification by delving into the aspects of a user's Twitter presence, that most reliably predict platform verification.

\subsection{Motivation}

Our motivation behind this work was two-fold and is elaborated in the following text.

\textbf{Lack of procedural clarity and imputation of bias:} Despite repeated statements by Twitter about verification not being equivalent to endorsement, aspects of the process -- the rarity of the status and its prominent visual signalling~\cite{b14} -- have led users to conflate authenticity and credibility. This perception was confirmed in full public view when Twitter was backed into suspending its requests for verification in response to being accused of granting verified status to political extremists~\footnote{\url{https://www.bbc.com/news/technology-41934831}}, with the insinuation being that the verified badge lent their otherwise extremist opinions a facade of mainstream credibility. 

This however, engendered accusations of Twitter's verification procedure harbouring a liberal bias. Multiple tweets imputing the same gave rise to the hashtag \#VerifiedHate. Similar insinuations have been made by right-leaning Indian users of the platform in the lead up to the 2019 Indian General Elections under the hashtag \#ProtestAgainstTwitter. These hitherto unfounded allegations of bias prompted us to delve deeper into understanding what may be driving the process and inferring whether these claims were justified or could the difference in status be explained away by less insidious factors relating to a user's profile and content. 

\textbf{Positive perception and coveted nature:} Despite having its detractors, the fact remains that a verified badge is highly coveted amongst public figures and influencers. This is with good reason as in spite of being intended as a mark of authenticity, prior work in social sciences and psychology points to verified badges conferring additional credibility to a handle's posted tweets~\cite{b4,b5,b6}. Psychological testing~\cite{b10} has also revealed that the credibility of a message and its reception is influenced by its purported source and presentation rather than just its pertinence or credulity. Captology studies~\cite{b11} indicate that widely endorsed information originating from a well-known source is easier to perceive as trustworthy and back up the former claim. This is pertinent as owners of verified accounts are usually well-known and their content is on an average more frequently liked and retweeted than that of the generic Twittersphere~\cite{b16,b15}.

Adding to the desirability of exclusive visual indicators is the demanding nature of credibility assessment on Twitter. The imposed character limit and a minimal scope of visually customizing content, coupled with the feverish rate at which content is consumed -- with users on average devoting a mere three seconds of attention per tweet~\cite{b13} -- makes users resort to heuristics to judge online content. There is substantial work on heuristic based models for online credibility evaluation~\cite{b7,b12,b55}. Particularly relevant to this inquiry is the \textit{endorsement heuristic}, which is associated with credibility conferred to it (e.g. a verified badge) and the \textit{consistency heuristic}, which stems from endorsements by several authorities (e.g. a user verified in one platform is likely to be verified on others).

Unsurprisingly, a verified status is highly sought after by preeminent entities, as evidenced by the prevalence of get-verified-quick schemes such as promoted tweets from the now suspended account `@verified845'~\cite{b18,b19}. Our work attempts to obtain actionable insights into verification process, thus providing entities looking to get verified a means to strengthen their case.

\subsection{Research Questions}

The aforementioned motivating factors pose a few avenues of research enquiry which we attempt to answer in this work are are detailed below.
\begin{itemize}
\item [\textbf{RQ1:}] Can the verification status of a user be predicted from profile metadata and tweet contents? If so what are the most reliably discriminative features?

\item [\textbf{RQ2:}] Do any inconsistencies exist between verified and non-verified users with respect to peripheral aspects like the choice and variety of topics they tweet about?
\end{itemize}

\subsection{Contributions}

Our contributions can be summarized as follows:
\begin{itemize}
    \item We motivate and propose the problem of predicting verification status of a Twitter user.
    \item We detail a framework extracting a substantial set of features from data and meta-data about social media users, including friends, tweet content and sentiment, activity time series, and profile trajectories. We plan to make this dataset of 407,165 users and 494 million tweets, publicly available upon publication of the work.\footnote{\url{http://precog.iiitd.edu.in/requester.php?dataset=twitterVerified19}}
    \item Additionally, we factored in state-of-the-art bot detection analysis into our predictive model. We use these features to train highly-accurate models capable of discerning a user's verified status. For a general user, we are able to provide a zero to one score representing their likelihood of being verified in Twitter.
    \item We report the most informative features in discriminating verified users from non-verified ones and also shed light on the manner in which the span and gamut of topic coverage between their tweets differs.
\end{itemize}

The rest of the paper is organized as follows. Section~\ref{sec:related} details relevant prior work, hence putting our work in perspective. Section~\ref{sec:dataset} elaborates our data acquisition methodology. In Sections~\ref{sec:resultsAndAnalysis} and~\ref{sec:topicAnalysis}, we conduct a comparative analysis between verified and non-verified users, addressing RQ1 and RQ2 respectively, and attempt to uncover features that can reliably classify them. We conclude with a brief summary in Section~\ref{sec:conclusions}. 

\section{Related Work}
\label{sec:related}
Previous studies have focused on measuring user impact in social networks. As user impact might be a critical factor in deciding who gets verified on Twitter~\cite{b1}, it is important to study how certain users in particular networks have more impact/influence as compared to the others. Cha et al.~\cite{cha2010measuring} studied the dynamics of influence on Twitter based on three key measures: in-degree, retweeets, and user-mentions. They show that in-degree alone is not sufficient to measure the influence of a user on Twitter. Bakshy et al.~\cite{bakshy2011everyone} demonstrate that URLs from users who have been influential in the past tend to generate larger cascades on the Twitter follower graph. They also show that URLs considered more interesting and that kindle positive emotions, spread more. Canali et al.~\cite{canali2012quantitative} identify key users on social networks who are important sources or targets for content disseminated online. They use a dimensionality-reduction based technique and conduct experiments with YouTube and Flickr datasets to obtain results which outperform the existing solutions by 15\%. The novelty of their approach is that they use attribute rich user profiles and not just stay limited to their network information. On the other hand, Lampos et al.~\cite{lampos2014predicting} predict user impact on Twitter using features, such as user statistics and tweet content, that are under the control of the user. They experiment with both linear and non-linear prediction techniques and find that Gaussian Processes based models perform the best for the prediction task. Klout~\cite{kloutservice} was a service that measured the influence of a person using information from multiple social networks. Their initial framework~\cite{rao2015klout} used long lasting (e.g., in-degree, pagerank centrality, recommendations etc) and dynamic features (reactions to a post such as retweets, upvotes etc.) to estimate the influence of a person across nine different social networks.

Further studies have tried to classify users based on factors such as celebrity status, socioeconomic status etc. Lampos et al.~\cite{lampos2016inferring} classify the socioeconomic status of users on Twitter as high, middle or lower socioeconomic, using features such as tweet content, topics of discussion, interaction behaviour, and user impact. They obtain an accuracy of 75\% using a nonlinear, generative learning approach with a composite Gaussian Process kernel. Preoctiuc-Pietro et al.~\cite{preoctiuc2015studying} present a Gaussian Process regression model, which predicts the income of the user on Twitter. They examined factors that help characterize user income on Twitter and analyze their relation with emotions, sentiments, perceived psycho-demographics, and language used in posts. Further, Marwick et al.~\cite{marwick2011see} qualitatively study the behaviours of celebrities on Twitter and how it impacts creation and sharing of content online. They aim to conceptualize ``celebrity as a practice'' in terms of personal information revelation, language usage, interactions, and affiliation with followers, among other things. There are also other studies that try to characterize usage patterns~\cite{al2015human} and personalities~\cite{tadesse2018personality} of varied users on Twitter.

Multiple existing studies attempt to detect and analyze automated activity on Twitter~\cite{chu2012detecting,zhang2011detecting,gilani2017depth,dickerson2014using,wang2010detecting,chavoshi2016identifying} and differentiate bot activity from human or partial-human activity. Conversely, Chu et al.~\cite{chu2012detecting} identify users on Twitter that generate automated content. The verification badge was a key feature used for the purpose. Holistically characterizing features that resemble automated activity, and the extent to which exhibiting the same can hurt a user's case for verification is further explored in Section~\ref{cluster}.

Past studies on verified accounts have focused on elucidating their behaviors and properties on Twitter. Hentschel et al.~\cite{hentschel2014finding} analyze verified users on Twitter and further use this information to identify trustworthy ``regular'' (not fake or spam) Twitter users. Castillo et al.~\cite{b4} attempt to identify credible tweets based on a variety of profile features including whether the user was authenticated by the platform or not. Along similar lines, Morris et al.~\cite{b5} examined factors that influence profile credibility perceptions on Twitter. They found that possessing an authenticated status is one of the most robust predictors of positive credibility. Paul et el.~\cite{b17} performed multiple network analyses of the verified accounts present on Twitter and reveal how they diverge from earlier results on the network as a whole. Hence, to summarize, there exists a rich body of literature establishing the enhancement of credibility and perceived importance a verified badge endows a user with. However, no prior work, to the best of our knowledge, has attempted to characterize attributes that make the aforementioned status more attainable.

\section{Dataset}
\label{sec:dataset}
In this section, we present details of our dataset and the data collection process along with a summary of the diverse features.

\subsection{User Metadata}

The \href{https://twitter.com/verified}{`@verified'} handle on Twitter follows all accounts on the platform that are currently verified. We queried this handle on the 18\textsuperscript{th} of July 2018 and extracted the IDs of 297,776 users (of which 231,235 have their primary language set to English) who were verified at the time. In the interest of verifying Twitter's assertion that likeliness of an handle's verification is commensurate with public interest in that handle and nothing else~\cite{b1,b30}, we sought to obtain a random controlled subset of non-verified users on the platform. Pursuant to this need, we leveraged Twitter's Firehose API -- a near real-time stream of public tweets and accompanying author metadata -- in order to acquire a random set of 284,312 non-verified users, controlling for a conventional measure of public interest, by ensuring that the number of followers of every non-verified user obtained was within 2\% that of a unique verified user that we had previously acquired.

Twitter provides a REST Application Programming Interface (API) with various endpoints that make data retrieval from the site in an organized manner easier. We used the REST API to acquire profile metadata of the user handles obtained previously including account age, number of friends, followers and tweets. Additionally, we obtained the number of public Twitter lists a user was part of and the handle's profile description. Metadata features extracted from user profiles have previously been used for classifying users and inferring activity patterns on Twitter~\cite{b24,b25}. We further focused our work to the subset of users who had English listed as their profile language thus enabling us to focus on the largest linguistic group on the platform~\cite{b31} and leaving us with 231,235 English verified users and 175,930 non-verified users.

\subsection{Content Features}

Utilizing Twitter's Firehose API, we acquired all tweets authored by the aforementioned users over a one year collection period spanning from 1\textsuperscript{st} June 2017 to 31\textsuperscript{st} May 2018. In total, our collection process acquired roughly 494,452,786 tweets. The tweet texts were retained and any accompanying media such as GIFs were deemed surplus to requirements and discarded.

From the text we extracted linguistic and stylistic features such as the number and proportion of \textit{Part-Of-Speech} (POS) tags, effectively obtaining a user's breakdown of natural language component usage. Work demonstrating the importance of content features in location inference~\cite{b33}, tweet classification~\cite{b37}, and network characterization~\cite{b35} further led us to extract the frequency of hashtags, retweets, mentions and external links used by each user. Prompted by studies showing that the deceptiveness of tweets could be inferred from the length of sentences constituting them~\cite{b32}, we computed additional features including average words per sentence, average words per tweet, character level entropy and frequency and proportion of long words (word length greater than six letters) per user.

In the interest of better discerning the emotions conveyed by the tweets authored by a user and responses they may evoke in the potential audience, sentiment analysis presented itself as an effective tool. Sentiment gleaned from Twitter conversations has been used to predict financial outcomes~\cite{b38}, electoral outcomes~\cite{b27} as well as the ease of content dissemination~\cite{b39}. We used Vader~\cite{b23}, a popular social media sentiment analysis lexicon, which has previously been widely used in a plethora of applications ranging from predicting elections~\cite{b27,b29} to forecasting cryptocurrency market fluctuations~\cite{b40}. We extracted positive, negative and neutral sentiment scores and an additional fourth compound score, which is a nonlinear normalized sum of valence computed based on established heuristics~\cite{b66} and a sentiment lexicon. All four scores are computed per user, weighted by tweet length.

\subsection{Temporal Features}

Existing research suggests that temporal features relating to content generation and activity levels on Twitter can be used to infer emergent trending topics~\cite{b41} as well as influential users~\cite{b42}.

\begin{table*}[t!]
\begin{threeparttable}
\centering
\begin{tabular}{ c l | c l }
\hline
 \parbox[t]{2mm}{\multirow{10}{*}{\rotatebox[origin=c]{90}{\textbf{User Metadata}}}} & Number of followers & \parbox[t]{2mm}{\multirow{10}{*}{\rotatebox[origin=c]{90}{\textbf{Temporal Features}}}}
 & Average number of followers last year \Tstrut\\ 
 & Number of friends & & Average number of friends last year \\
 & Number of statuses & & Average number of statuses last year \\
 & Number of public list memberships & & Proportion of followers gained in last 3 months \\
 & Account age & & Proportion of friends gained in last 3 months \\
 & & & Proportion of statuses generated in last 3 months \\
 & & & Proportion of followers gained in last 1 month \\
 & & & Proportion of friends gained in last 1 month \\
 & & & Proportion of statuses generated in last 1 month \\
 & & & Average duration between statuses \Bstrut\\
 \hline
  \parbox[t]{2mm}{\multirow{14}{*}{\rotatebox[origin=c]{90}{\textbf{Content Features}}}} & Number of POS tags\tnote{1} & \parbox[t]{2mm}{\multirow{14}{*}{\rotatebox[origin=c]{90}{\textbf{Miscellaneous Features}}}} & LIWC analytic summary score \Tstrut\\ 
 & Frequency of POS tags\tnote{1} & & LIWC authentic summary score  \\  
 & Average number of words per sentence & & LIWC clout summary score \\
 & Average number of words per tweet & & LIWC tone summary score \\
 & Character level entropy & & Botometer complete automation probability \\
 & Proportion of long words\tnote{2} & & Botometer network score \\
 & Positive sentiment score\tnote{3} & & Botometer content score \\
 & Negative sentiment score\tnote{3} & & Botometer temporal score \\
 & Neutral sentiment score\tnote{3} & & Tweet topic distribution\tnote{4} \\
 & Compound sentiment score\tnote{3} & &  \\
 & Frequency of hashtags & &  \\
 & Frequency of retweets & &  \\
 & Frequency of mentions & &  \\
 & Frequency of external links posted & &  \Bstrut\\
 \hline
\end{tabular}
\caption{List of features extracted per user by our framework.}\label{tab:1}
\begin{tablenotes}
     \item[1] Part Of Speech (POS) tags include nouns, personal pronouns, impersonal pronouns, adjectives, adverbs, verbs, auxiliary verbs, prepositions and articles.
     \item[2] Long words are defined as words longer than 6 letters.
     \item[3] Sentiment scores are weighted over all tweets of a user by tweet length.
     \item[4] Scores over 100 topics are extracted from the tweets.
   \end{tablenotes}
  \end{threeparttable}
\end{table*}

Leveraging the Twitter Firehose, we gathered fine-grained time series of user statistics including number of friends, followers and statuses, thus permitting us to compute their averages over our one year collection period. Furthermore, positing that a user's likelihood of verification may be predicated on how ascendant their reach in the platform is, we compute the proportion of friends and followers gained over the last one month and the last three months of our collection period. Additionally, similar trajectory encoding features are computed for tweet activity levels over the aforementioned one and three month windows, and the average time between statuses is extracted using the status count time series on a per user basis.

\subsection{Miscellaneous Features}

Attempting to capture qualitative cognitive and emotional cues from a user's tweets, we acquired the four LIWC 2015~\cite{b46} summary statistics named Analytic, Clout, Authentic and Tone for each user in our dataset. The summary dimensions indicate the presence of logical and hierarchical thinking patterns, confidence and leadership, personal cues and emotional tone, respectively, in the tweets of a user. LIWC categories have been scientifically validated to perform well in determining affect on Twitter~\cite{b49,b26} and have been previously used to detect sarcasm~\cite{b48} and for mental health diagnoses from Twitter conversations~\cite{b47}.

Furthermore, positing that accounts perceived as being completely or partially automated may have a harder time getting verified, we leveraged Botometer -- a flagship bot detection solution~\cite{b20} that exposes a free public API. The system is trained on thousands of instances of social bots and the creators report AUC ROC scores between 0.89 and 0.95. Botometer utilizes features spanning the gamut from network attributes to temporal activity patterns. Additionally, it queries Twitter to extract 300 recent tweets and publicly available account metadata, and feeds these features to an ensemble of machine learning classifiers, which produce a Complete Automation Probability (CAP) score, which we acquire for every user in our dataset. We also augment our dataset with the temporal, network and content category automation scores for each user.

Finally, we also look to glean into the topics that users tweet about. Topic modelling has been effectively used in categorizing trending topics on Twitter~\cite{b36} and inferring author attributes from tweet content~\cite{b28}. To this end, we ran the Gibbs sampling based Mallet implementation of Latent Dirichlet Allocation (LDA)~\cite{b50} setting the number of topics to 100 with 1000 iterations of sampling. Although, such a topic model could be applied on a per tweet basis and subsequently aggregated by user, we find this approach to not work very well as most tweets are simply a sentence long. To overcome this difficulty, we follow the workaround adopted by previous studies by aggregating all
the tweets of a user into a single document~\cite{b44,b45}. In effect, this treatment can be regarded as an application of the author-topic model~\cite{b43} to tweets, where each document has a single author.

\subsection{Rectifying Class Imbalance}

Focusing our analysis on the Twitter Anglosphere left us with a substantially skewed class distribution of 231,235 verified users and 175,930 non-verified users in our dataset. In keeping with existing research on imbalanced learning on Twitter data~\cite{b51,b52}, we used a two-pronged approach to rectify this -- a minority over-sampling technique named ADASYN~\cite{b21} which generates samples based on the feature space of the minority examples and a hybrid over and under-sampling technique called SMOTETomek which additionally also eliminates samples of the over-represented class~\cite{b22} and has been found to give exemplary results on imbalanced datasets\cite{b53}. Augmenting our classifier's training data in the aforementioned manner allowed us to attain near-perfect classification scores.

The data collected is classified and summarized in Table~\ref{tab:1}. We intend to anonymize and make this dataset accessible to the public in a manner compliant with Twitter terms, once this work is published.

\section{Results and Analysis}
\label{sec:resultsAndAnalysis}
We commence our analysis by eliminating all features that could be deemed surfeit to requirements. To this end, we employed an all-relevant feature selection model~\cite{b58} which classifies features into three categories: confirmed, tentative and rejected. We only retain features that the model is able to confirm over 100 iterations.

To evaluate the effectiveness of our framework in discerning verification status of users, we examine five classification performance metrics -- precision, recall, F1-score, accuracy and area under ROC curve -- for five classifiers. The first two methods intended at establishing baselines were a Logistic Regressor and a Support Vector Classifier. Further, three methods were used to gauge how far the classification performance could be pushed using the features we collected. These were (1) a Generalized Additive Model trained by nested iterations, setting all terms to smooth, (2) a Multi Layered Perceptron with 3 hidden layers of 100, 30 and 10 neurons respectively, using Adam as an optimiser and ReLU as activation and (3) state-of-the-art Gradient Boosting tool named XGBoost with a maximum tree depth of 6 and a learning rate of 0.2.  The results obtained are detailed in Table~\ref{tab:2}. The first batch of results are obtained by training on the original unadulterated training split. Even without rectifying class distribution biases, we are able to attain a high classification accuracy of 98.9\% on our most competitive classifier.

The second and third batches are trained on data rectified for class imbalance using the adaptive synthetic over-sampling method (ADASYN) and a hybrid over and under-sampling method (SMOTETomek), respectively. The ADASYN algorithm generates samples based on the feature space of the minority class data points and is a powerful method that has seen success across many domains~\cite{b59} in neutralizing the deleterious effects of class imbalance. The SMOTETomek algorithm combines the above over-sampling strategy with an under-sampling method called Tomek link removal~\cite{b60} to remove any bias introduced by over-sampling. This rectification did improve results, generally improving the performance of our two baseline choices and especially helping us inch closer to perfect performance with gradient boosting. However, particularly surprising was the detrimental effect of class re-balancing on the MLP classifier which in all likeliness also learned the non-salient patterns in the re-balanced data. Also unexpectedly, the ADASYN re-balancing outperformed the more sophisticated SMOTETomek re-balancing in pushing the performance limits of the support vector (89.1\% accuracy) and gradient boosting (99.1\% accuracy) approaches. This might be owing to the fact that the Tomek link removal method omits informative samples close to the classification boundary thus affecting the learned support vectors and decision tree splits.

Our results suggest that near perfect classification of the Twitter user verification status is possible without resorting to complex deep-learning pipelines that sacrifice interpretability.

\begin{table*}[t!]
\begin{threeparttable}
\centering
\begin{tabular}{ | c | l | c | c | c | c | c | }
\hline
\textbf{Dataset} & \textbf{Classifier} & \textbf{Precision} & \textbf{Recall} & \textbf{F1-Score} & \textbf{Accuracy} & \textbf{ROC AUC Score}\Tstrut\Bstrut\\
\hline
 &Logistic Regression &0.86 &0.86 &0.86 &0.859 &0.854\Tstrut\\
Original &Support Vector Classifier &0.89 &0.89 &0.89 &0.887 &0.883 \\
imbalanced &Generalized Additive Model\tnote{1} &0.97 &0.98 &0.98 &0.975 &0.976 \\
data &3-Hidden layer NN (100,30,10) ReLU+Adam &0.98 &0.98 &0.98 &0.983 &0.977 \\
 &XGBoost Classifier &0.99 &0.99 &0.99 &\textbf{0.989} &\textbf{0.990}\Bstrut\\
\hline
 &Logistic Regression &0.86 &0.86 &0.86 &0.856 &0.858\Tstrut\\
ADASYN &Support Vector Classifier &0.89 &0.89 &0.89 &0.891 &0.891 \\
class &Generalized Additive Model\tnote{1} &0.97 &0.97 &0.97 &0.974 &0.973 \\
rebalancing &3-Hidden layer NN (100,30,10) ReLU+Adam &0.96 &0.96 &0.96 &0.959 &0.957 \\
 &XGBoost Classifier&0.99 &0.99 &0.99 &\textbf{0.991} &\textbf{0.991}\Bstrut\\
\hline
 &Logistic Regression &0.86 &0.86 &0.86 &0.860 &0.856\Tstrut\\
SMOTETomek &Support Vector Classifier &0.90 &0.90 &0.90 &0.903 &0.901 \\
class &Generalized Additive Model\tnote{1} &0.98 &0.97 &0.98 &0.974 &0.974 \\
rebalancing &3-Hidden layer NN (100,30,10) ReLU+Adam &0.97 &0.97 &0.97 &0.966 &0.968 \\
 &XGBoost Classifier&0.99 &0.99 &0.99 &\textbf{0.990} &\textbf{0.991}\Bstrut\\
\hline
\end{tabular}
\caption{Summary of classification performance of various approaches using metadata, temporal and contextual features on the original and balanced datasets.}\label{tab:2}
\begin{tablenotes}
    \item[1] The generalized additive models were trained using all smooth terms.
\end{tablenotes}
\end{threeparttable}
\end{table*}

\subsection{Feature Importance Analysis}

\begin{figure}[!hbt]
\centerline{\includegraphics[width=0.95\linewidth,height=9.5cm]{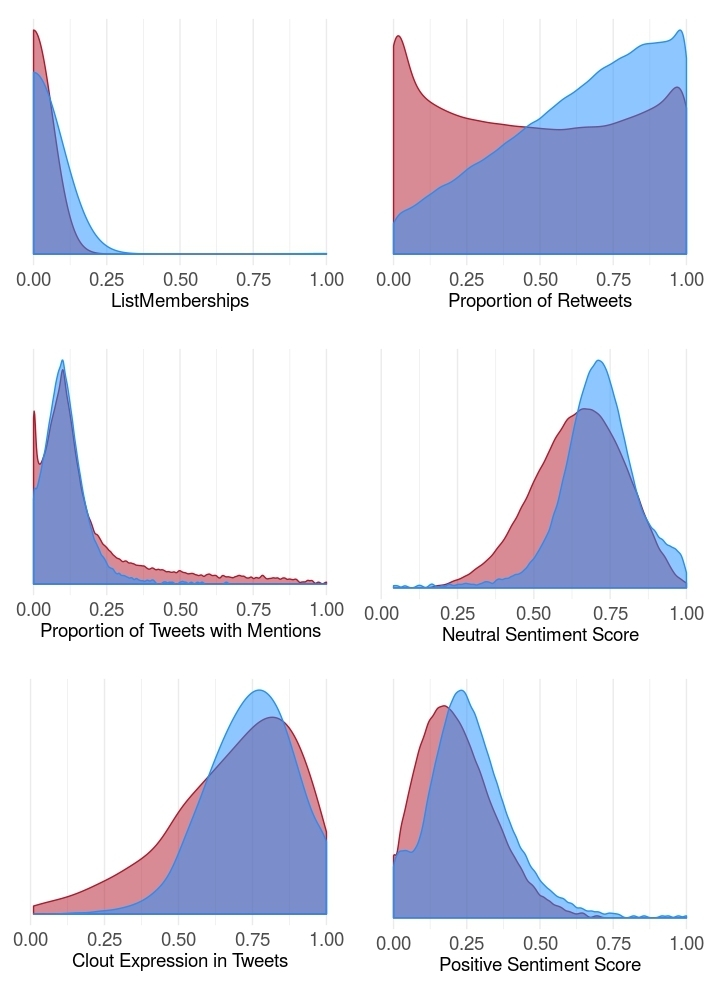}}
\caption{\textbf{Normalized density estimations of the six most discriminative features for verified (blue) and non-verified users (red).}}
\label{fig:fig1}
\end{figure}

To compare the usefulness of various categories of features, we trained gradient boosting classifier, our most competitive model, using each category of features alone. While we achieved the best performance with user metadata features, content features were not far behind. Evaluated on multiple randomized train-test splits of our dataset, user metadata and content features were both able to consistently surpass 0.88 AUC. Additionally, temporal features alone are able to consistently attain an AUC of over 0.79.

The individual feature importances were determined using the Gini impurity reduction metric output by the gradient boosting model trained on the unmodified dataset. To rank the most important features reliably, the model was trained 100 times with varying combinations of hyperparameters (column sub-sampling, data sub-sampling and tree child weight) and the features determined to be the most important were noted. The most reliably discriminative features and their normalized density distributions over the values they attain are detailed in Figure~\ref{fig:fig1}. These features generally exhibit intuitive patterns of separation based on which an informed prediction can be attempted, e.g., the very highest echelons of public list membership counts are populated exclusively by verified users while the very low extremes of propensity for authoritative speech as indicated by LIWC Clout summary scores are exclusively displayed by non-verified users. 

The top 6 features are sufficient to reach performance of 0.9 AUC on their own right and the top 10 features are sufficient to further push those numbers up to 0.93. This is largely owing to the fact that substantial redundancy was observed among sets of highly correlated features such as some linguistic (tendency to use long words and impersonal pronouns highly correlate with high analytic LIWC summary scores) and temporal trajectory (most ascendant users score highly in both the 1 month and 3 month features in terms of tweets authored and followers gained) features.

\subsection{Clustering and characterization}
\label{cluster}

\begin{figure}[!hbt]
\centerline{\includegraphics[width=0.95\linewidth,height=6cm]{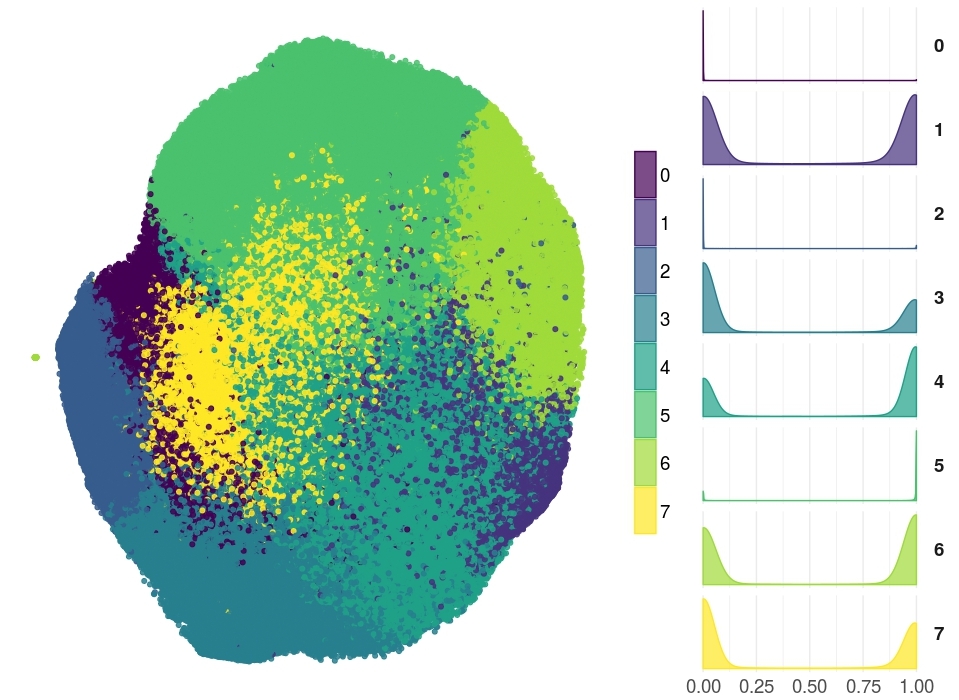}}
\caption{\textbf{t-SNE embeddings of accounts coloured by cluster. The distribution of verification probabilities by cluster, as predicted by our classifier, are faceted on the right.}}
\label{fig:fig2}
\end{figure}

In order to characterize accounts with a higher resolution than a binary verification status will permit, we apply K-Means++ on the normalized user vectors selecting the 30 most discriminative features indicated by the XGBoost model -- our most competitive classifier. We settle on 8 different clusters based on evaluation including the inflection point of the clustering inertia curve and the proportion of variance explained. In the interest of an intuitive visualization, two dimensional embeddings obtained using t-SNE dimensionality reduction method~\cite{b54} are presented. Tuning the perplexity metric appropriately, the method considers the similarity of data points in our feature space and embeds them in a manner that reflects their proximity in the feature space. The embeddings are plotted and our classifier responses for members of the different clusters are detailed in Figure~\ref{fig:fig2}. 

Investigating these clusters allows us to further unravel combinations of attributes that strengthen a user's case for verification. Clusters C0 and C2 are composed nearly exclusively of non-verified users. Cluster C0 can largely be characterized as the Twitter layman with a high proportion of experiential tweets. This narrative further plays out in our collected features with members of this cluster on average having short tweets, high incidence of verb usage and scoring especially high in the LIWC Authenticity summary. Cluster C2 can be characterized as an amalgamation of accounts exhibiting bot-like behavior. Members of this cluster scored highly on the complete, network and content automation scores in our feature set. Furthermore, members in C2 possessed attributes previously linked to spammers such as copious usage of hashtags~\cite{b56} and external links~\cite{b57}. Manual inspection verified the substantial presence of automated content such as local weather updates in this cluster. Unsurprisingly, members of this cluster were predicted to possess the lowest verification probability by our classifier.

The composition of clusters C4 and C6 leans towards verified users, with members of C4 having a tendency to post longer tweets and retweet more frequently than author content, while members of C6 almost exclusively retweet on the platform with slightly over 93\% of their content being such. Cluster C5 is nearly entirely comprised of verified users and includes elite Twitteratti that comprise the core of verified users on the platform. These users have by far the highest list memberships on average while also scoring very highly on the LIWC Clout summary. Predictably, members of this cluster were predicted to possess the highest verification probability by our classifier.

\begin{table}[ht!]
\begin{threeparttable}
\centering
\begin{tabular}{ | c | c | c | c | }
\hline
\textbf{Cluster} & \textbf{Population} & \textbf{Accuracy} & \textbf{ROC AUC Score}\Tstrut\Bstrut\\
\hline
C0 & 19462 & 0.996 & 0.989\Tstrut\\
C1 & 26259 & 0.986 & 0.986\\
C2 & 19356 & 0.994 & 0.984\\
C3 & 46178 & 0.988 & 0.987\\
C4 & 90843 & 0.989 & 0.987\\
C5 & 105701 & 0.993 & 0.986\\
C6 & 39248 & 0.990 & 0.989\\
C7 & 60118 & 0.987 & 0.986\Bstrut\\
\hline
\end{tabular}
\caption{Classification performance of our most competitive model broken down by cluster.}\label{tab:3}
\end{threeparttable}
\end{table}

The remaining clusters C1, C3 and C7 are comprised of a mix of verified and non-verified users. However, further inspection revealed that they have very divergent trajectories. Members of cluster C1 are ascendant both in terms of reach and activity levels as evidenced by the proportion of their followers gained and statuses authored in the last one and three months of our collection period. These members can be said to constitute a nouveau-elite group of users. This is further backed up by the fact that these users are lacking in their presence in public lists as compared to the very established elite in cluster C5. Manual inspection also verifies that many of these users have attained verification during our collection period. This is in stark contrast with members of C3 and C7 who are either stagnant or declining in their reach and activity levels and show very low engagement with the rest of the platform in terms of retweets and mentions. Remarkably, our classifier is able to make this distinction and rates members of C1 as slightly better candidates for verification on average than members of C3 or C7. The relative difficulty of classifying users in these mixed clusters is demonstrated in the performance breakdown detailed in Table ~\ref{tab:3}.

\section{Topic Analysis for Verified vs Non-Verified Users}
\label{sec:topicAnalysis}

Having deduced important predictive features present in a user's metadata, linguistic style and activity levels over time with respect to verification status, we next investigate the presence of similar predictive patterns in the choice and variety of tweet topic usage amongst users. 

\subsection{Content Topics}

\begin{figure}[!bht]
\centerline{\includegraphics[width=0.95\linewidth,height=9.5cm]{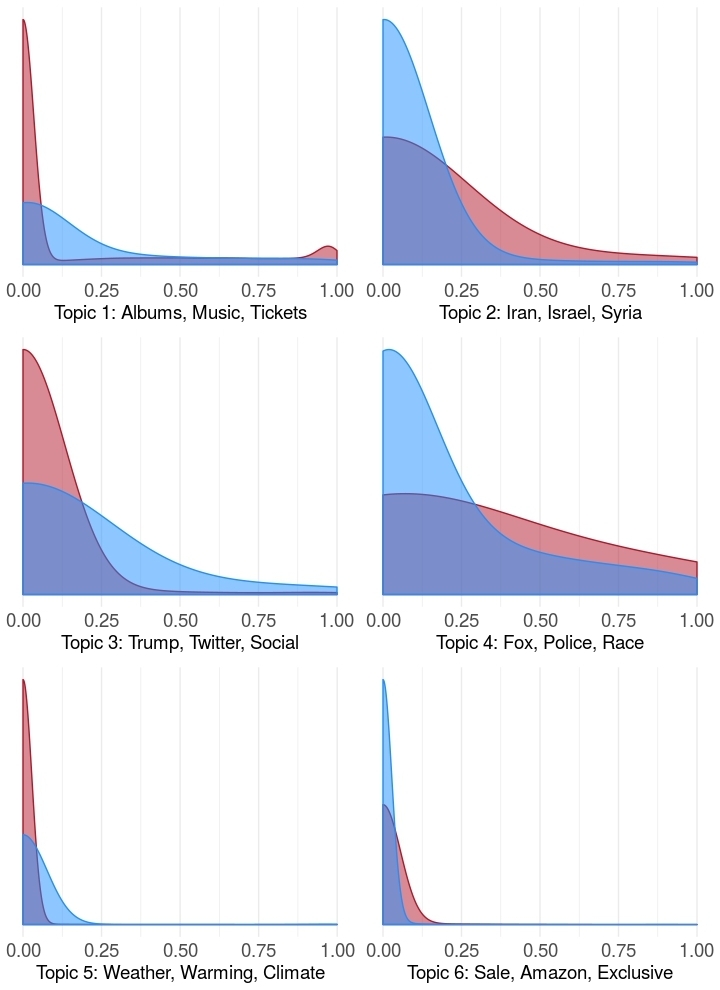}}
\caption{\textbf{Normalized density estimations of usage for the six most discriminative topics for verified (blue) and non-verified users (red). Listed alongside are the top three most probable keywords for each topic.}}
\label{fig:fig3}
\end{figure}

\begin{table*}[t!]
\begin{threeparttable}
\centering
\begin{tabular}{ | r | c | c | c | c | c | }
\hline
\textbf{Classifier} & \textbf{Precision} & \textbf{Recall} & \textbf{F1-Score} & \textbf{Accuracy} & \textbf{ROC AUC Score}\Tstrut\Bstrut\\
\hline
Generalized Additive Model\tnote{1} &0.83 &0.83 &0.83 &0.832 &0.831 \Tstrut\\
3-Hidden layer NN (100,30,10) ReLU+Adam &0.88 &0.88 &0.88 &\textbf{0.882} &\textbf{0.880} \\
XGBoost Classifier &0.82 &0.82 &0.82 &0.824 &0.823 \Bstrut\\
\hline
\end{tabular}
\caption{Summary of classification performance of various approaches on inferred topics.}\label{tab:4}
\begin{tablenotes}
\item[1] The generalized additive models were trained using all smooth terms.
\end{tablenotes}
\end{threeparttable}
\end{table*}

In order to obtain a topical breakdown of a user's tweets in an unsupervised manner, we ran the Gibbs sampling based Mallet implementation of Latent Dirichlet Allocation (LDA)~\cite{b50} with  1000 iterations of sampling. Narrowing down on the correct number of topics $T$ required us to execute multiple runs of the model while varying our choices for the number of topics. The model was executed for 30, 50, 100, 150 and 300 topics and the likelihood estimates were noted. It must be mentioned that in all cases the likelihood estimates stabilized well within the 1000 iteration limit we set. The likelihood keeps rising in value up to $T = 100$ topics, after which it sees a decline. This kind of profile is often seen when varying the hyperparameter of a statistical model, with the optimal model being rich enough to fit the information available in the data, yet not complex enough to begin fitting noise. This led us to conclude that the tweets we collected over a year are best accounted for by incorporating 100 separate topics. We set $\alpha = T/50 $ and $\beta = 0.01 $, which are the default settings recommended in prior studies~\cite{b62} and maintain the sum of the Dirichlet hyperparameters, which can be interpreted as the number of virtual samples contributing to the smoothing of the topic distribution, as constant. The chosen value of $\beta$ is small enough to permit a fine-grained breakdown of tweet topics covering various conversational areas.

We again commenced the prediction by pruning down our topical feature set using the all relevant feature selection method we used earlier~\cite{b58} in Section ~\ref{sec:resultsAndAnalysis}. This allowed us to hone in on the 76 topics that were confirmed to be predictive of verification status. To evaluate the effectiveness of our framework in discerning verification status of users from topic cues, we examine five classification performance metrics -- precision, recall, F1-score, accuracy and area under ROC curve -- for the three classifiers that were most competitive in our previous classification task. These were (1) a Generalized Additive Model trained by nested iterations, setting all terms to smooth, (2) a Multi Layered Perceptron with 3 hidden layers of 100, 30 and 10 neurons respectively, using Adam as an optimiser and ReLU as activation and (3) Gradient Boosting tool named XGBoost with a maximum tree depth of 5 and a learning rate of 0.3. The results obtained are detailed in Table \ref{tab:4}. The results demonstrate that it is eminently possible to infer the verification status of a user purely using the distribution of topics they tweet about with a high accuracy. The MLP classifier was the most competitive in this task, reliably pushing past 88.2\% accuracy. 

In the interest of interpretability, we evaluate the predictive power of each topic with respect to the classification target. To this end, we obtain individual topic importances using the ANOVA F-Scores output by GAM -- our second most competitive model on this task. In order to rank the features reliably, the procedure is run on 50 random train-test splits of the dataset and the topics with the lowest F-Scores noted. The most reliably discriminative topics and the normalized density distributions of their usage are detailed in Figure~\ref{fig:fig3}. Owing to multiple topics largely belonging to popular broad conversational categories such as sports and politics, some redundancy was observed in the way of multi-collinearity. This is further backed up by the fact that the top 15 most important topics alone can discern verification status with an AUC of 0.76 while the top 25 topics can push those numbers up to an AUC of 0.8 nearly approximating the GAM performance on the whole feature set (AUC 0.83). These topics generally exhibit intuitive patterns of separation based on which an informed prediction can be made, e.g., the users who tweet most frequently about climate change are all verified while controversial topics like middle-east geopolitics are something verified users prefer to devote limited attention to.

\subsection{Topical Span}

\begin{figure}[!bht]
\centerline{\includegraphics[width=0.95\linewidth,height=6cm]{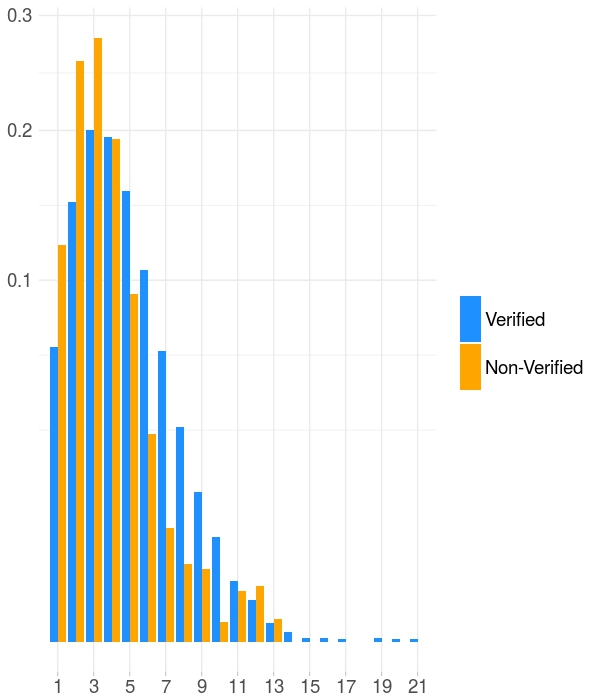}}
\caption{\textbf{Square-root scaled proportion of users by optimal number of topics.}}
\label{fig:fig4}
\end{figure}

Peripheral aspects of topics such as their geographical distribution~\cite{b64} and the viability of embeddings they induce for sentiment analysis~\cite{b64} tasks have been explored before. This prompted us to extend our inquiry into peripheral measures such as inconsistencies in the variety and number of topics the two classes of users tweet about. In order to obtain an optimal mix of the number of topics per user in an unsupervised manner, we leveraged the use of an Hierarchical Dirichlet Process (HDP) model implementation~\cite{b63} for topic inference. This method streams our corpus of tweets and performs an online Variational Bayes estimation to converge at an optimal number of topics $T$, for each user. Once again, we set $\alpha = T/50 $ and $\beta = 0.01 $, which are the default settings recommended in existing studies~\cite{b62}.

The distribution of cardinality for topic sets by verification status are detailed in Figure~\ref{tab:4}. Inspection of the distribution uncovers a clear trend with non-verified users clearly being over-represented in the lower reaches of the distribution (1--4 topics), while a comparatively substantial portion of verified users are situated in the middle of the distribution (5--10 topics). Also noteworthy is the fact that the very upper echelons of topical variety in tweets are occupied solely by verified users. We posit that this may be owing to the fact that news handles (e.g., \href{https://twitter.com/BBC}{`@BBC'}: 13 topics) and content aggregators (e.g.,  \href{https://twitter.com/gifs}{`@GIFs'}: 21 topics) are over represented in the set of verified users. The validation of this assertion is left for future work. 

\section{Conclusion}
\label{sec:conclusions}

The coveted nature of platform verification on Twitter has led to the proliferation of verification scams and accusations of systemic bias against certain ideological demographics. Our work attempts to uncover actionable intelligence on the inner workings of the verification system, effectively formulating a checklist of profile attributes a user can work to improve upon to render verification more attainable.

This article presents a framework that computes the strength of a user's case for verification of Twitter. We introduce our machine learning system that extracts a multitude of features per user, belonging to different classes: user metadata, tweet content, temporal signatures, expressed sentiment, automation probabilities and preferred topics. We also categorize the users in our dataset into intuitive clusters and detail the reasons behind their likely divergent outcomes from the verification procedure. Additionally, we demonstrate role, that a user's choices and variety over conversational topics plays in precluding or effecting verification.

Our framework represents the first of its kind attempt at discerning and characterizing verification worthy users on Twitter and is able to attain a near perfect classification performance of 99.1\% AUC. We believe this framework will empower the average Twitter user to significantly enhance the quality and reach of their online presence without resorting to prohibitively priced social media management solutions.

\bibliographystyle{ACM-Reference-Format}
\balance
\bibliography{sample-base}

\end{document}